  \documentstyle[graphicx,prl,aps,epsf,twocolumn]{revtex}

  \title{Electric instability in superconductor-normal conductor ring}
  \author{A. Kadigrobov \cite{e-mail}
  }
  \address{ Theoretische Physik III, Ruhr-Universitt Bochum, D-44780 Bochum, Germany}

\begin{document}

  \date{}
  \maketitle
  \begin{abstract}

Non-linear electrodynamics of a ring-shaped Andreev interferometer (superconductor-normal conductor-superconductor hybrid structure) inductively coupled to a circuit of the dissipative current is investigated. The current-voltage characteristics (CVC) is
demonstrated to be  a series of loops with  several branches  intersecting  in the CVC  origin. The sensitivity of the transport current  to a change of the applied external   magnetic flux can be comparable to the one of  the conventional SQUID's. Spontaneous arising of coupled non-linear  oscillations of the transport current,
 the Josephson current  and the magnetic flux  in Andreev interferometers are also predicted and investigated. The frequency of these oscillations can be varied in a wide range, while the maximal frequency can reach $\omega_{max} \sim 10^{12}$ $sec^{-1}$.
  \end{abstract}

  Recent years much attention has been paid to the charge transport in mesoscopic systems
  which combine  normal conductor (N) and superconductor (S) elements (for  review papers see, e.g., \cite{Lambert,Bruder} and references there).
  In such a hybrid structure superconducting correlations penetrate into the normal conductor
  within mesoscopic distances changing its transport properties there.
  This quantum effect is most pronounced in  S-N-S structures ("Andreev interferometers") in which the quantum interference gives rise to a high sensitivity of these
  superconducting correlations  to the phase difference
  between the superconductors $\varphi$.
  It results, in particular, in oscillations of  the Josephson current $J_s=J_s(\varphi)$ with a change of  $\varphi$, the amplitude of the oscillations \cite{Bruder}
  being $J_c \sim N_\perp e v_F/L_N$ ($N_\perp = S_N/\lambda_F^2$,  $\lambda_F=\hbar/p_F$ - the Fermi wave length, $p_F$ - the electron Fermi momentum, $e$ - the electron charge, $v_F$ - the Fermi velocity, $S_N$ and $L_N$ are the cross-section area and the length of the normal conductor,  respectively).

  If the Andreev interferometer is coupled to reservoirs of normal electrons,
  the conductance $G(\varphi)$ and hence the  dissipative current, $J_d=G(\varphi)V $, between the reservoirs
  ($V$ is the bias voltage) are also phase-dependent,
  the amplitude of the conductance oscillations being proportional to $N_\perp e^2/\hbar$ (in many experiments
  $N_\perp \sim 10^2\div 10^5$).

\begin{figure}
 \centerline{\includegraphics[width=8.0cm]{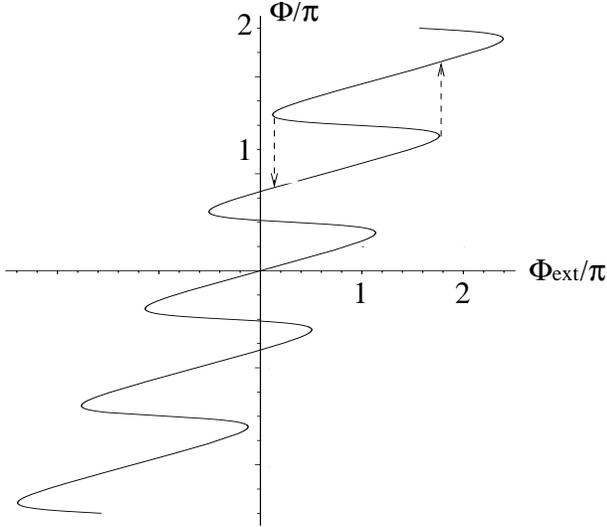}}
  \vspace*{2mm}
  \caption{Dependence of the magnetic flux threading  the ring, $\Phi $, on the applied  flux
    $\Phi_{ext}=H_{ext} S_r. $}
\label{cvcflux1}
  \end{figure}

  In a ring-shaped geometry of the S-N-S structure,  the phase
  difference between the superconductors is controlled by the magnetic flux $\Phi= H S_r$ ($H$ is the magnetic field threading the ring, $S_r$ is the area of the  ring):
  $\varphi = -2\pi \Phi /\Phi_0$
where the flux quantum $\Phi_0 = \pi \hbar c/e$, $c$ -the light velocity. This allows to control  the Josephson and  the dissipative currents with a change of the external magnetic field $H_{ext}$
that is favorable for various applications. On the other hand, the   Josephson current (which depends on the magnetic flux in the ring)
creates its proper magnetic field which
 modifies this  flux. As a result, for a ring of a large enough self-inductance,  the relation between  the magnetic flux  $\Phi$ and the external
magnetic field $H_{ext}$ turned out to be highly non-linear  with hysteresis loops in the dependence of  $\Phi$ on  $H_{ext}$   as shown in Fig.\ref{cvcflux1} (see, e.g.,  \cite{flux}). This self-inductive property
of the Josephson current has been used for creation of Superconducting Quantum Interference Devices (SQUID) which have found
extensive applications  for  extremely precise measurements \cite{flux}.
    \begin{figure}
 \centerline{\includegraphics[width=8.0cm]{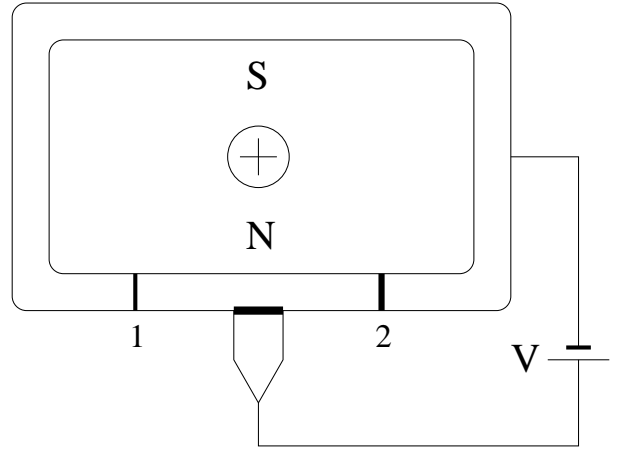}}
  \vspace*{2mm}
  \caption{Superconductor-normal conductor-superconductor structure of the Andreev interferometer type to which a voltage drop $V$ is applied. Thick lines indicate potential barriers at  normal conductor - superconductor interfaces 1 and 2, and between the lead and the normal section of the interferometer.}
\label{sample1}
  \end{figure}

A dissipative  current flow through the S-N-S ring creates a new situation in which
 the Josephson and  dissipative
currents are coupled together  through their joint influence on the magnetic flux inside the S-N-S ring that, in its turn, affects the both currents themselves.

\begin{figure}
  \centerline{\includegraphics[width=8.0cm]{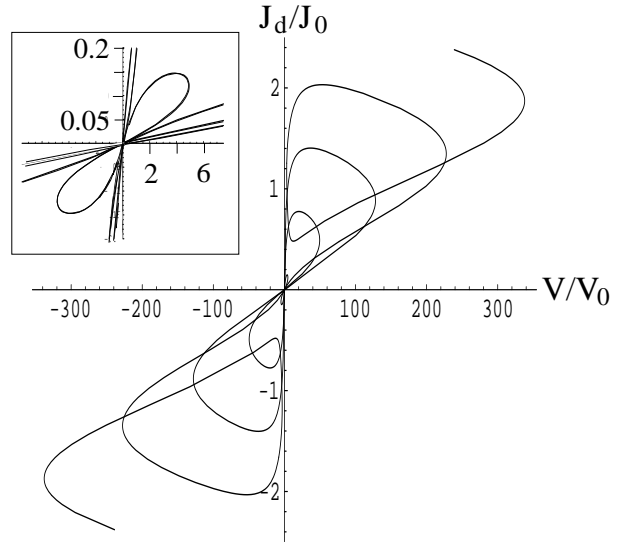}}
  \vspace*{2mm}
  \caption{Current - voltage characteristics of an SNS ring for the external flux
   $\Phi_{ext}/\Phi_0=(2l+1)\pi $, $l=0, \pm 1,\pm 2, ...$; $t_r =0.1$, $t_N^{(1)}=0.2$, $t_N^{(2)}=0.25$, and ${\cal L}_r = 2.5 c \Phi_0/J_c$; $J_0 = ({\cal L}_r /{\cal L}_{12} )J_c$, $V_0=2 \hbar v_F/(\pi e L_N)$. The insertion shows a zoomed  vicinity of the CVC origin.}
  \label{cvc1}
  \end{figure}

The aim of this paper is to show that  the inductive interaction between the dissipative and Josephson currents
results in a complicated loop-shaped form
  of the current-voltage characteristics (CVC) of S-N-S structures schematically presented in Fig.\ref{sample1}. In the general case
several branches of the CVC (which correspond to different values of the magnetic flux $\Phi$ inside the ring) intersect  in the origin  of CVC ($J_d =0$, $V=0$) as is shown in the insertion of  Fig.(\ref{cvc1}).
Stable  non-linear periodic in time oscillations of the dissipative current $J_{d}$, the Josephson current $J_s$ and the magnetic flux $\Phi$
 inside the ring (and hence the phase difference  $\varphi$) are predicted and investigated.  The form of the loop-shaped
CVC and the frequency of  these oscillations  are shown to be extremely sensitive to the applied external magnetic
field $H_{ext}$  and the  bias voltage $V$.

The system under consideration is a ring-shaped superconductor which has a  section of a normal conductor  (see Fig.\ref{sample1}).
A steady bias voltage $V$ is applied between the normal and the superconducting sections of the ring that provides a dissipative
current $J_d$ in this circuit. There is  a potential barrier
 of a low  transparency $t_r$ between the normal conductor and the lead. The  electron transport may be in the ballistic or in the diffusive regimes.
1. {\it Loop-shaped current-voltage characteristics}. The total flux of the magnetic field threading the ring, $\Phi$,
is created by the external magnetic field $H_{ext}$, the proper magnetic fields of the Josephson current $J_s$ and the dissipative current $J_d$, and hence can be written as
\begin{equation}
  \Phi =\Phi_{ext}+\frac{{\cal{L}}_{r}}{c}J_s+\frac{{\cal{L}}_{12}}{c}J_d
  \label{totalflux}
  \end{equation}
where $\Phi_{ext}=H_{ext} S_r$ is the magnetic flux of the external magnetic field $H_{ext}$, ${\cal{L}}_{r}$
is the self-inductance of the ring, ${\cal{L}}_{12}$ is the mutual inductance of the ring and the dissipative current circuit.

Eq.(\ref{totalflux}) together with the conventional relation  $J_d = G(\varphi)V$
define a parametric form of the CVC:
\begin{eqnarray}
J_d =J_d (\varphi)\equiv \frac{1}{{\cal L}_{12}}\left(\frac{c \Phi_0}{2\pi}(\varphi_{ext}-\varphi)-{\cal{L}}_{r}J_s(\varphi)\right)
  \nonumber \\
 V= \frac{J_d(\varphi)}{G(\varphi)}; \hspace{1 cm} \varphi_{ext}\equiv -2\pi \Phi_{ext}/\Phi_0  \hspace{1.5 cm}
\label{cvc}
  \end{eqnarray}

 One of the  distinguishing  features of the CVC of the system under consideration is its "many-fold degeneracy",  that is  several branches of the CVC
(which correspond to different values of the magnetic flux threading the ring $\Phi$) can  intersect
in its origin ($V=0$, $J_d=0$). As is seen from Eq.(\ref{cvc}) at $J_d=0$ and Fig.\ref{cvcflux1}, the "degeneracy factor" is equal to the number
of intersections of the vertical line $\Phi_{ext}= const$ and  the curve $\Phi=\Phi(\Phi_{ext})$.

Other  key features of the CVC can be found
if one consider the differential resistance $dV/dJ_d$ which is readily obtained from Eq.(\ref{cvc}):
\begin{eqnarray}
  \frac{d V}{dJ_d} = \frac{1}{G(\varphi)}\left(1 + \frac{2\pi }{c \Phi_0}\frac{{\cal{L}}_{12}J_d}{G(\varphi)}\frac{ G^\prime(\varphi)}{ A(\varphi)}
  \right)_{\varphi = \overline{\varphi}(J_d)}
 \label{diffresistance}
  \end{eqnarray}
Here and below the sign "prime" means the derivative with respect to $\varphi$;  the function $A(\varphi)$ \cite{L12} is
\begin{eqnarray}
A(\varphi)= 1+\frac{2\pi {\cal{L}}_r}{c \Phi_0}J_s^\prime(\varphi)
 \label{A}
  \end{eqnarray}
All quantities in Eq.(\ref{diffresistance}) are taken
at  $\varphi =\overline{\varphi}(J_d)$ which is a solution of the first equation
in Eq.(\ref{cvc}). Using Eq.(\ref{cvc}) one finds
\begin{eqnarray}
d\overline{\varphi}/dJ_d = -\frac{2\pi {\cal{L}}_{12}}{c \Phi_0}\frac{1}{A(\overline{\varphi})}
 \label{deriv}
  \end{eqnarray}

If   ${\cal{L}}_r >{\cal{L}}_{cr}^{(1)} \equiv  c \Phi_0/(2\pi max\{|J_s^\prime|\}) $ there are values of $\varphi$
 at which $A(\varphi)=0$.
Therefore,  in this case,
as follows  from Eq.(\ref{diffresistance}) the CVC $J_d(V)$ inevitably has points at which  the differential resistance
$d V /dJ_d$  goes to infinity changing its sign there.
On the other hand, the derivative
$G^\prime$ also changes its sign with a change of $\varphi$
 that can provide points at which
 $d V /dJ_d$ goes to zero changing  its sign at them. The consequtive order of this peculiarities with an increase of $J_d$
 (that is the form of the CVC) depends on the relative positions of the maxima and minima of $G(\varphi)$ and  $J_s(\varphi)$.

\begin{figure}
 \centerline{\includegraphics[width=8.0cm]{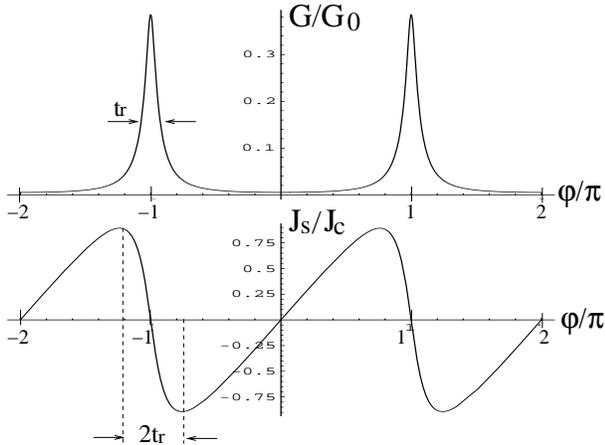}}
 \vspace*{2mm}
  \caption{Typical dependences of the Josephson current and the conductance on the superconductor phase difference $\varphi$
  for the case of a low transparency ($t_r \ll 1$) of the potential barrier between the normal section of the sample and the lead
  ($J_c= N_\perp e v_F/L_N$ and $G_0=N_\perp e^2/(\pi \hbar)$). }
\label{joscond}
  \end{figure}

For the case of a low transparency of the potential barrier  $t_r \ll 1$ the conductance $G(\varphi)$ has maxima at odd numbers of $\pi$   \cite{Lambert} while the Josephson current $J_s =0$ at them and has a maximum and a minimum in their vicinities \cite{Bruder} as schematically shown in Fig.\ref{joscond}.
This general information together with Eq.(\ref{diffresistance} - \ref{deriv}) would suffice to find the CVC to be loop-shaped if ${\cal{L}}_r >{\cal{L}}_{cr}^{(1)}$. In order to see it let us start from the  external magnetic field which corresponds to $\overline{\varphi}=0$ and hence $G^\prime =0$, $J^\prime >0$ and therefore $dV/dJ_d > 0$ according to Eq.(\ref{diffresistance}). As at this point $d\overline{\varphi}/d J_d <0$  the phase difference $\overline{\varphi}$ decreases with an increase of  current $J_d$ (see Eq.(\ref{deriv})). Therefore, with an increase of the current $G^\prime$ becomes negative and  after passing the minimum
of $J_s(\overline{\varphi})$, $A(\overline{\varphi})$ starts to decrease because  $J^\prime <0$ now (see Fig.\ref{joscond}), and when $A$ is small enough (but  positive yet), the differential resistance becomes $dV/dJ_d = 0$. With a further
increase of the current $A(\overline{\varphi})\rightarrow +0$  while $G^\prime $ is yet negative, and hence $dV/dJ_d \rightarrow -\infty$.
When $A(\overline{\varphi}) = -0$ the differential resistance $dV/dJ_d = +\infty$. In order to follow this second branch of CVC one should decrease
the  current because $d \overline{\varphi} /d J_d >0$ (see Eq.(\ref{deriv})). Pursuing such a reasoning one easily finds the current-voltage characteristic  to be a series of loops which  touch  the lines $J_d =G_{min}V$ and $J_d =G_{max}V$
(in our case $G_{min} = G(0)$, $G_{max} = G(\pi)$); the number of loops intersecting in the origin of the CVC increases with an increase
of the ring inductance. Such a loop-shaped CVC takes place for the both cases of a diffusive or a ballistic normal conductor of the
S-N-S ring-shaped interferometer.

An example of such a  current-voltage characteristics is presented in Fig.\ref{cvc1} for the case of a ballistic normal section
  of the Andreev interferometer in the presence of  potential barriers
  at the N/S interfaces. In this case the  dependences of the conductance $G(\varphi)$ \cite{GCO} and the Josephson current $J_s(\varphi)$ can be written  as follows :

\begin{eqnarray}
G(\varphi) =
\frac{t_r^2 G_0}{\sqrt{\left(1 + |r_A^{(1)} r_A^{(2)}|\cos{\varphi} +t_r^2/2\right)^2 - |r_N^{(1)}r_N^{(2)}|^2}} \label{conductance}\\
J_s(\varphi) =
N_\perp\frac{2ev_F^{\|}}{\pi L_N} |r_A^{(1)}r_A^{(2)}|\sin{\varphi} \times \hspace{2.8 cm}\nonumber\\
\int_0^{2 \pi}  \frac{d\phi}{2\pi \sin{\varphi_{-}(\phi)}}\arctan{
\frac{e^{-2t_r}\sin{\varphi_{-}(\phi)}}{1-e^{-2t_r}\cos{\varphi_{-}(\phi)}}}
 \label{Josephson}
  \end{eqnarray}
Here $G_0=N_\perp e^2/(\pi \hbar)$, $v_F^{\|}=N_\perp^{-1}\sum_{\vec{n}}^{N_\perp }v_{\vec{n}} \sim v_F$ and $v_{\vec{n}}$ is the electron velocity in the $\vec{n}$-th transverse mode; $r_N^{(1,2)}$ are the normal  reflection amplitudes at   N/S interfaces 1 and 2 while $r_A^2+r_N^2=1$, (see \cite{Blonder}); $\cos{\varphi_{-}(\phi)}=|r_N^{(1)}r_N^{(2)}|\cos{\phi}-|r_A^{(1)}r_A^{(2)}|\cos{\varphi}$ and $\sin{\varphi_{-}(\phi)}=\sqrt{1-\cos^2{\varphi_{-}(\phi)}}$. Expanding  Eq.(\ref{Josephson}) in $\exp\{\varphi_{-}\}$ and taking the limit $t_r \rightarrow 0$, $t_N^{(1,2)}\rightarrow 0$ one  reduces it to the well-known expression for the Josephson current in a 3D SNS junction \cite{Svidzinski}.

   Inserting  Eq.(\ref{conductance},\ref{Josephson}) in Eq.(\ref{cvc}) and performing numerical calculations
one obtains the current-voltage characteristics shown in Fig.\ref{cvc1}.

   2. {\it Electromagnetic self-oscillations.} A change of the magnetic flux threading a superconductor-normal conductor ring results in the following  current in the ring:
 \begin{eqnarray}
J_r = J_s(\varphi)-\frac{1}{cR}\frac{d \Phi}{d t}
 \label{Jr}
  \end{eqnarray}
where $R$ is the resistance of the normal section of the S-N-S ring \cite{Jr}. Using  $\varphi = -2\pi \Phi /\Phi_0$ one sees Eq.(\ref{Jr}) to be the equation of the ac Josephson effect.

 Taking into account the self-inductances of the ring ${\cal{L}}_r$ and the transport current circuit ${\cal{L}}_{11}$ together with their mutual inductances ${\cal{L}}_{12}$, after  using Eq.(\ref{totalflux}) (in which $J_s$ should be changed to $J_r$) and Eq.(\ref{Jr})  one gets a set of equations that describes the time evolution
of the transport current and the superconductor phase difference as follows:
\begin{eqnarray}
\frac{{\cal{L}}_{11}}{c^2}\frac{d J_d}{d t}+\frac{{\cal{L}}_{12}\Phi_0}{2\pi c^3R}\frac{d^2\varphi}{d t^2}
+ \frac{{\cal L}_{12}J_s^{\prime}(\varphi)}{c^2 }\frac{d\varphi}{d t}+ \frac{J_d}{G(\varphi)}=V;\nonumber\\
\frac{\Phi_0{\cal{L}}_r}{ 2\pi c R}\frac{d\varphi}{d t} +\frac{c \Phi_0}{2\pi}\varphi+  {\cal L}_r J_s(\varphi) +{\cal L}_{12}J_d=-c \Phi_{ext}
\hspace{0.7 cm}
\label{nonsteady1}
  \end{eqnarray}
The static solutions of Eq.(\ref{nonsteady1}) give the  CVC (Eq.(\ref{cvc})).

In order to investigate time-evolution of the system it is convenient to eliminate $J_d$ from the set of equations Eq.(\ref{nonsteady1}) and
get the following closed equation for $\varphi(t)$:
\begin{eqnarray}
\frac{{\cal L}_{eff}}{Rc^2}\frac{d^2 \varphi}{dt^2}+\gamma (\varphi)\frac{d\varphi}{d t}+F(\varphi)=0
 \label{phaseeq}
  \end{eqnarray}
where  ${\cal L}_{eff}=$ $({\cal L}_r{\cal L}_{11} - {\cal L}_{12}^2)/{\cal L}_{11})$. Eq.(\ref{phaseeq}) is the equation for a non-linear oscillator under a "friction"

 \begin{eqnarray}
\gamma(\varphi)=1+\frac{2\pi{\cal L}_{eff}}{c\Phi_0}J_s^{\prime}(\varphi)+\frac{{\cal L}_r}{{\cal L} _{11}}\frac{1}{R G(\varphi)}
 \label{friction}
  \end{eqnarray}
and a "force"
 \begin{eqnarray}
F(\varphi)= \frac{c^2}{{\cal L} _{11} G(\varphi)}
\times  \hspace{3.4cm} \nonumber \\
\left(\varphi -\varphi_{ext}+\frac{2\pi{\cal L}_r}{c\Phi_0}J_s(\varphi)+\frac{2\pi{\cal L}_{12}}{c\Phi_0}G(\varphi)V\right)
 \label{force}
  \end{eqnarray}

For low values of ${\cal L}_{11}$ the "friction" $\gamma>0$ at any value of $\varphi$ and hence, starting from any initial state, the system approaches one of its static states determined by Eq.(\ref{totalflux}) (in which $J_d=G(\varphi)V$). With an increase of ${\cal L}_{11}$,
the "friction" $\gamma(\varphi)$ becomes  negative in a certain range of $\varphi$ and the static state can become unstable. Investigations of the stability of the static solutions of Eq.(\ref{phaseeq}) $\varphi=\varphi_{st}$ show that
the critical value of ${\cal L}_{11}$ is determined by the condition $\gamma (\varphi_{st})=0$, that is
\begin{eqnarray}
{\cal L}_{cr}=\frac{{\cal L}_r}{B RG(\varphi_{st})};  \hspace{2.3 cm}\nonumber \\B= -\left(1+\frac{{2\pi\cal L}_{eff}}{c\Phi_0}J_s^\prime(\varphi_{st})\right)>0
 \label{critic}
  \end{eqnarray}
The Poincare method \cite{Poincare} shows
that in the plane $(\varphi, \dot \varphi)$ a stable limit cycle
 arises if
 $$L_{11}>{\cal L}_{cr}; \hspace{0.2 cm} b \equiv 1+\frac{2 \pi {\cal L}_r}{c \Phi_0} J_s^\prime (\varphi_{st})+ \frac{2 \pi {\cal L}_{12}}{c \Phi_0}  G^\prime (\varphi_{st})>0 $$
These inequalities (together with the one in Eq.(\ref{critic})) can be satisfied if only $J_s^\prime (\varphi_{st})<0$ and $G^\prime (\varphi_{st})>0$. From here and Eq.(\ref{diffresistance},\ref{A}) it follows that the stable limit cycle (that is non-linear periodic time-oscillations of $\varphi(t)$ and $\dot \varphi(t)$) can only be on those  branches of  the CVC with  the negative differential resistance $d V/DJ_d <0$ that are close to the $J_d$-axis in Fig.\ref{cvc} (and on the loop shown in the insertion of the figure). If $0< {\cal L}_{11}-{\cal L}_{cr} \ll {\cal L}_{cr}$ the frequency of these   oscillations is
$$\omega_0=\sqrt{\frac{b c^4 R}{\left({\cal L}_{11}{ \cal L}_r - {\cal L}_{12}^2 \right)G(\varphi_{st})}} $$
This frequency can be variated in a  wide range, and  estimations show the maximal
 frequency  $\omega_{max}\sim 10^{12}$ $sec^{-1}$.

In conclusion, for both the diffusive and ballistic cases  I have shown  that   the current-voltage characteristics of the Andreev ring-shaped interferometer $J_d (V)$ is a series of loops with
 several branches intersecting in its  origin, and hence it
 has sections  with a negative differential resistance $d V /dJ_d$  if the ring inductance is large enough.
These properties of the CVC  are robust as  the negative differential resistance and
the intersection of the several branches in the CVC origin  appear at any value of the
mutual inductance, ${\cal L}_{12}$, between the ring and the dissipative transport circuit (which can be  as small as is wished)  as soon as  the self-inductance of the ring ${\cal L}_r$ is large enough to provide $A<  0$ (see Eq.(\ref{diffresistance},\ref{A}); of course, such a CVC can not be observed experimentally if the mutual inductance  ${\cal L}_{12}$ is unreasonably small.
 This inequality ($A<  0$) is the same as the one needed  for the conventional SQUID functioning.  What's is more, manipulations with the ring self-inductance ${\cal L}_r$ and the external magnetic flux $\Phi_{ext}$ allow to get
 CVC loops (and therefore the hysteresis loops of the CVC) as small as wanted, especially it concerns  the smallest loop around the CVC origin (see Insertion in Fig.\ref{cvc1}), and  one can obtain the sensitivity of the transport current $J_d$ (or the applied voltage $V$) to a change of the external magnetic flux comparable  to the conventional SQUID sensitivity.

  Spontaneous arising of coupled non-linear  oscillations of the transport current $J_d(t)$,
 the Josephson current $J_s(t)$ and the  flux $\Phi(t)$  in Andreev interferometers have been also predicted and investigated. The frequency of the oscillations $\omega$ can be varied in a wide range, and  the maximal frequency  can reaches $ \omega_{max} \sim 10^{12}$ $sec^{-1}$.


\begin{thebibliography}{99}
  \bibitem{e-mail} e-mail: kadig@tp3.ruhr-uni-bochum.de
  \bibitem{Lambert} C.J. lambert and R. Raimondi, J. Phys.: Condens. Matter {\bf 10}, 901 (1998).
  \bibitem{Bruder} C. Bruder, Supercond. Rev. {\bf 1}, 261 (1996).
  \bibitem{flux} M. Cyrot and D. Pavuna, "Introduction to Superconductivity and High-T$_c$ Materials", Scientific World  1992.
    \bibitem{L12}
    In general, ${\cal{L}}_{12}$   depends on the distribution of  $J_d$ inside the SNS ring which, as shown in
    paper \cite{Shumeiko1},  depends on $\varphi$. However, we neglect this dependence of ${\cal{L}}_{12}$ on $\varphi$ as it  can only renormalize ${\cal L}_{cr}^{(1)}$ (see the text under Eq.(\ref{deriv})), to say nothing of the event that the transport current circuit is a solenoid with the S-N-S ring inside (or around) it. In this case the influence of the current distribution inside the ring on ${\cal{L}}_{12}$ is completely negligible.
    \bibitem{Shumeiko1} P. Samuelsson, V.S. Shumeiko, and G. Wendin, Phys. Rev.B {\bf 56}, R5763 (1997);
    P. Samuelsson, J. Lantz, V.S. Shumeiko, and G. Wendin, Phys. Rev.B {\bf 62}, 1319 (2000).
    \bibitem{GCO}  H.A. Blom,
     A. Kadigrobov, A.M. Zagoskin, R.I. Shekhter, and M. Jonson,
    Phys. Rev.B {\bf 57}, 9995 (1998).
     \bibitem{Blonder}
    A. L. Shelankov, JETP Lett. {\bf 32}, 111 (1980); G. E. Blonder, M. Tinkham, and T. M. Klapwijk,  Phys.
    Rev.B {\bf 25}, 4515 (1982).
     \bibitem{Svidzinski} C. Ishii, Prog. Theor. Phys. {\bf 44}, 1525 (1970); J. Bardin and J.L. Johnson, Phys. Rev. B {\bf 5}, 72 (1972);
     A.V. Svidzinski, T.N. Anzygina, and E.N. Bratus, Sov. Phys. JETP {\bf 34}, 860 (1972).
    \bibitem{Jr} This equation is readily obtained if one uses the  equation $J_r =J_s(\varphi) + (L_N/R)\int \{(- 1/c) d\vec{A}/dt - \nabla v\}d\vec{l}$ for the normal section and the London equation for  the supercurrent density in the superconducting section.
 ($\vec{A}$ is the vector potential, $v$ is the voltage in the ring).
    \bibitem{Poincare} A.A. Andronov, A.A. Witt, S.E. Khaikin,  "Theory of Oscillators", Oxford: Pergamon 1966.
  \end{thebibliography}
    \end{document}